\newcounter{myctr}
\def\myitem{\refstepcounter{myctr}\bibfont\noindent\ifnum\themyctr>9\else\phantom{0}\fi\hangindent17pt\themyctr.\enskip}

\documentclass{ws-ijqi}
\usepackage{hyperref}
\usepackage[super,sort,compress]{cite}

\newcommand{\ket}[1]{\left|{#1}\right>}

\newcommand{\rvec}[1]{\pmb{#1}}
\newcommand{\dyadic}[1]{\pmb{#1}}

\newcommand{\D}{\mathrm{d}}
\newcommand{\I}{\mathrm{i}}
\newcommand{\TP}[1]{{#1}^\mathrm{\,\textsc{t}}}
\newcommand{\E}[1]{\mathrm{e}^{\mbox{\footnotesize$#1$}}}
\newcommand{\Tr}[1]{\mathrm{Tr}\!\left\{#1\right\}}

\newcommand{\DET}[1]{\det\!\left\{#1\right\}}

\newcommand{\VAR}[2]{\mathrm{Var}_{#1}\!\left[#2\right]}
\newcommand{\MEAN}[2]{\mathbb{E}_{#1}\!\left[#2\right]}

\newcommand{\FHOMone}{\dyadic{F}_{1,\textsc{hom}}}
\newcommand{\FHOMtwo}{\dyadic{F}_{2,\textsc{hom}}}

\newcommand{\VEC}[1]{\mathrm{vec}\!\left(#1\right)}
\newcommand{\WEYLORD}[1]{\left[#1\right]_\textsc{ws}}
\newcommand{\sCRB}{\mathrm{sCRB}}
\newcommand{\MSE}{\mathrm{MSE}}
\newcommand{\rest}{\widehat{\rvec{r}}}
\newcommand{\Gest}{\widehat{\dyadic{G}}_2}

\begin{document}

\catchline{}{}{}{}{}

\title{JOINT MEASUREMENT OF COMPLEMENTARY OBSERVABLES IN MOMENT
	TOMOGRAPHY}

\author{YONG SIAH TEO}

\address{BK21 Frontier Physics Research Division, 
	Seoul National University, 08826 Seoul, South Korea\\
ys\_teo@snu.ac.kr}

\author{CHRISTIAN~R.~M\"{U}LLER}

\address{Max-Planck-Institut f\"ur  die Physik des Lichts,
	Staudtstra\ss e 2, 91058 Erlangen, Germany\\
christian.mueller@mpl.mpg.de}

\author{HYUNSEOK JEONG}
\address{Center for Macroscopic Quantum Control,
	Seoul National University, 08826 Seoul, South Korea\\
jeongh@snu.ac.kr }

\author{ZDEN{\v E}K HRADIL}
\address{Department of Optics, Palack\'{y}  University,
	17. listopadu 12, 77146 Olomouc, Czech Republic\\
hradil@optics.upol.cz}

\author{JAROSLAV \v{R}EH\'{A}\v{C}EK}
\address{Department of Optics, Palack\'{y}  University,
	17. listopadu 12, 77146 Olomouc, Czech Republic\\
rehacek@optics.upol.cz }

\author{LUIS L. S\'{A}NCHEZ-SOTO}
\address{Departamento de \'Optica, Facultad de F\'{\i}sica,
	Universidad Complutense, 28040 Madrid, Spain\\
lsanchez@fis.ucm.es}
\address{Max-Planck-Institut f\"ur  die Physik des Lichts,
	Staudtstra\ss e 2, 91058 Erlangen, Germany}

\maketitle

\begin{history}
\received{Day Month Year}
\revised{Day Month Year}
\end{history}

\begin{abstract}
Wigner and Husimi quasi-distributions, owing to their functional regularity, give the two archetypal and equivalent representations of all observable-parameters in continuous-variable quantum information. Balanced homodyning and heterodyning that correspond to their associated sampling procedures, on the other hand, fare very differently concerning their state or parameter reconstruction accuracies. We present a general theory of a now-known fact that heterodyning can be tomographically more powerful than balanced homodyning to many interesting classes of single-mode quantum states, and discuss the treatment for two-mode sources.
\end{abstract}



\markboth{Y. S. Teo et al}
{Joint measurement of complementary observables in moment tomography}

\section{Introduction}	

The successful implementation of any quantum-information protocol hinges on
the operational reliability of its individual components, which 
includes quantum sources that supply the resources for information 
transmission. Accurate calibrations of these sources are hence 
important and quantum tomography provides the necessary tools for 
this purpose. 

In continuous-variable quantum information theory, 
the technique of balanced homodyning (HOM)~\cite{Yuen:1983ba,Abbas:1983ak, Schumaker:1984qm,Vogel:1989zr,Banaszek:1997ot} samples the marginal distribution of the Wigner function of a given unknown state with approximate quadrature eigenstates. Heterodyning (HET)~\cite{Arthurs:1965al,Yuen:1982hh,Arthurs:1988aa,Martens:1990al,Martens:1991aa,Raymer:1994aj,Trifonov:2001up,Werner:2004as} on the other hand performs a delocalized sampling of the Husimi function of the state by a joint measurement of the complementary position $X$ and momentum $P$ operators---double-HOM so to speak. These two sampling methods probe the phase space in essentially two feasible ways to reconstruct observable parameters, that is the parameter column $\rvec{q}=\left<\rvec{V}\right>$ that depends on some state-independent observable column $\rvec{V}$: through either a direct sampling of some positive quasi-distribution or sampling physical aspects of some otherwise non-positive (and non-singular) quasi-distribution. 

In Refs.~\cite{Rehacek:2015qp} and \cite{Muller:2016da}, we showed that despite the fact that both Husimi and Wigner representations are mutually equivalent for describing quantum states, the reconstruction accuracies of observable parameters for schemes that probe these different quasi-distributions can be very different. In particular, we showed that for Gaussian states of a wide range of temperature $\mu\geq1$ (mean thermal photon number) and squeezing strength $\lambda\geq1$, HET beats HOM tomographically in reconstructing first and second moments even in the presence of additional vacuum noise originating from the joint measurement of complementary observables. The studies were based on the analysis of the optimal mean squared error, or the scaled Cram{\'e}r--Rao bound (sCRB). These results refuted a myth that suggests that because of the vacuum noise, ``\emph{the two }($X$-$P$)\emph{ beams measured have to suffer losses}'', such that ``\emph{each quadrature measurement will have a reduced SNR} (signal-to-noise ratio)''. An erroneously unsystematic assessment of the two schemes such as this would inevitably conclude with the belief that there ``\emph{is no advantage}'' in using HET~\footnote{These comments, which were directly extracted from a referee report for a journal that is not cited here, are representative of many statements for this myth.}, which is in direct contradiction with well-established experimental schemes~\cite{Muller:2016da,Croal:2016qs}.

In subsequent discussions, we shall present a general theory~\cite{Teo:2017aa} for the two sampling methods that applies to arbitrary single-mode states and extend this theory to two-mode states. We show that for a majority of the cases, HET beats HOM tomographically in terms of the sCRB. As examples, we investigate the performance of these methods in first and second-moment tomography on these states. The paper is organized as follows. We first state the general theory for moment tomography as well as for HOM and HET in Sec.~\ref{sec:gen_theory}. We then proceed to discuss first-moment and second-moment tomography for single-mode quantum states with more detail in Secs.~\ref{sec:first-mom} and \ref{sec:sec-mom}. More specifically, we prove a general optimality property of HET for first-moment tomography that holds for \emph{all} states, and in Secs.~\ref{subsec:second_mom_gauss} through \ref{subsec:second_mom_dispfock_padd}, we analyze the tomographic powers of HOM and HET defined by their sCRBs for interesting classes of states: the Gaussian, Fock, even and odd coherent, displaced Fock, and photon-added coherent states. Finally, we extend our discussions to two-mode states in Sec.~\ref{sec:two-mode}, taking a next step towards a more complete study of these schemes on general multimode photonic sources. There, we shall analyze two classes of two-mode states, the two-mode Fock and two-mode squeezed vacuum states.

\section{General theory}
\label{sec:gen_theory}

\subsection{Moments and tomographic power}
\label{subsec:mom_power}

In quantum mechanics, an arbitrary single-mode state $\rho$ can be characterized with an infinite set of operator moments, which are functions of the position $X$ and momentum $P$ operators~\cite{Englert:QM2} that parametrize the infinite-dimensional Hilbert space. When $\rho$ is a Gaussian state (a state with all quasi-distributions Gaussian), the Hilbert space is effectively parametrized by only the first and second moments. This turns the Hilbert space into a five-dimensional parameter space that is characterized by $\left<X\right>$, $\left<P\right>$, $\left<X^2\right>$, $\left<P^2\right>$ and $\left<\WEYLORD{XP}\right>$ where $\WEYLORD{XP}=(XP+PX)/2$ refers to the Weyl symmetrically-ordered operator moments. In this article, we study the performance of HOM and HET on the first and second moments for states besides the Gaussian ones. The corresponding results can be useful in many areas of quantum information theory, such as the topics of generalized uncertainty relations~\cite{Angulo:1993aa,Angulo:1994ps}, non-classicality detection~\cite{Simon:9709030,Arvind:1998aa}, entanglement detection~\cite{Namiki:2012gs,Ivan:2012da}, and cryptography~\cite{Leverrier:2012qs,Thearle:2016aa}.

It is convenient to group the first and second moments into the following two multivariate quantities:
\begin{align}
\rvec{r}&\,\widehat{=}\begin{pmatrix}
\left<X\right>\\
\left<P\right>
\end{pmatrix}\,,\quad\dyadic{G}_1=\rvec{r}\rvec{r}\,,\nonumber\\
\dyadic{G}_2&\,\widehat{=}\begin{pmatrix}
\left<X^2\right> & \left<\WEYLORD{XP}\right>\\
\left<\WEYLORD{XP}\right> & \left<P^2\right>
\end{pmatrix}\,.
\end{align}
The complete covariance matrix of any single-mode $\rho$ is then defined as $\dyadic{G}=\dyadic{G}_2-\dyadic{G}_1$, and contains all first- and second-moment information about $\rho$. Additionally, $\dyadic{G}$ satisfies the matrix inequality $\dyadic{G}\geq-\I\,\dyadic{\Omega}/2$ in terms of $\dyadic{\Omega}\,\widehat{=}\begin{pmatrix}
0 & 1\\
-1 & 0
\end{pmatrix}$ that is related to the two-dimensional symplectic group, or equivalently $\DET{\dyadic{G}}\geq1/4$, which is a consequence of the Heisenberg-Robertson-Schr{\"o}dinger (HRS) uncertainty relation.

The assessment of the reconstruction accuracies in reconstructing $\rvec{r}$ and $\dyadic{G}_2$ may be made more precisely by considering their mean squared-errors (MSE) $\MSE_1=\MEAN{}{(\rest-\rvec{r})^2}$ and $\MSE_2=\MEAN{}{\left(\Gest-\dyadic{G}_2\right)^2}$ for the respective estimators $\rest$ and $\Gest$. This tomographic measure encodes three kinds of information, namely the measurement performance, reconstruction strategy used to define the estimator and data sample size $N$ that is usually fixed by the observer for the sampling techniques of interest to us. In order to make a fair and conservative comparison between two different measurement performances, we define the so-called tomographic power to be the MSE that is minimized over all possible reconstruction strategies and scaled with $N$---the scaled Cram{\'e}r--Rao bound (sCRB). The expressions are given by $\sCRB_1=N\min_{\rest}\MEAN{}{(\rest-\rvec{r})^2}$ and $\sCRB_2=N\min_{\Gest}\MEAN{}{\left(\Gest-\dyadic{G}_2\right)^2}$.

\subsection{Balanced homodyning}

\begin{figure}[h!]
	\centering
	\includegraphics[width=1\columnwidth]{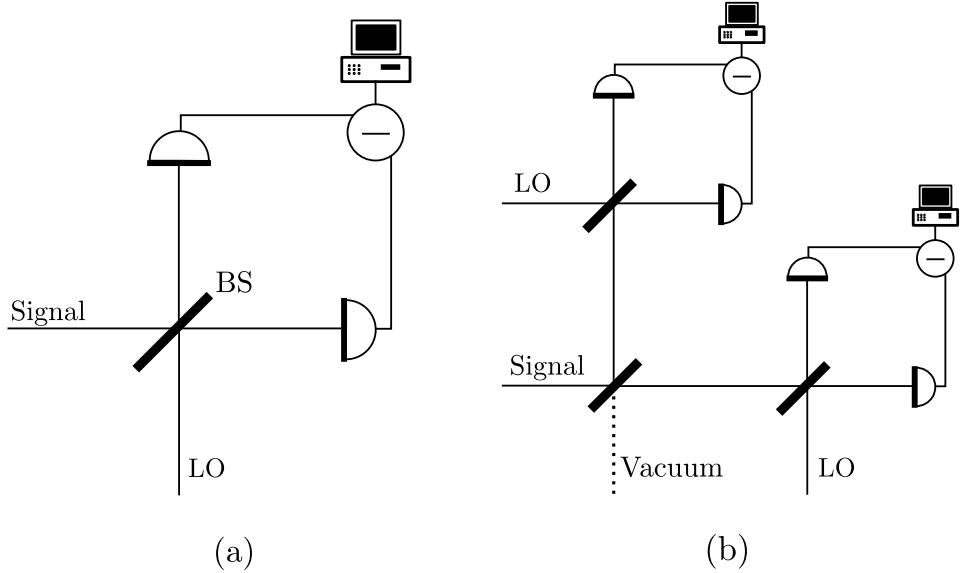}
	\caption{\label{fig:Fig1}Schema for the (a)~HOM and (b)~HET schemes. Here, BS denotes a 1:1 beam splitter and LO denotes the (strong) local oscillator.}
\end{figure}

The HOM scheme involves the coherent mixture of the optical signal and a strong local oscillator of complex amplitude $\alpha=|\alpha|\E{\I\vartheta}$ with a balanced (1:1) beam splitter, after which the two output photocurrents are subtracted and recorded as the real value $-\infty<x_\vartheta<\infty$ for a fixed phase $0\leq\vartheta<\pi$ [see~Fig.~\ref{fig:Fig1}(a)].

The distribution of $x_\vartheta$ is the marginal of the Wigner function along $\vartheta$, which measures the resulting quadrature observable $X_\vartheta=X\cos\vartheta+P\sin\vartheta$. To optimally reconstruct the moments with the HOM data, we take advantage of the simple formula
\begin{equation}
\left<\WEYLORD{X^mP^{M-m}}\right>=\left.\dfrac{1}{M!}\left(\dfrac{\partial}{\partial t}\right)^m\left(\dfrac{\partial}{\partial t'}\right)^{M-m}\left<(X t+P t')^{M}\right>\right|_{t=t'=0}
\end{equation}
that relates the Weyl ordered moments of order $M$ to the moments of $X_\vartheta$ ($0\leq m\leq M$), which states that all information about moments of this order is contained in $\left<X_\vartheta^{M}\right>$. Then the sCRB in moment tomography can be derived by calculating the Fisher information matrix $\dyadic{F}$ that defines it in the famous matrix-trace relation $\sCRB=N\Tr{\dyadic{F}^{-1}}$. In the limit of large $N$, the Fisher matrix for the $M$th moments can be shown to be
\begin{equation}
\dyadic{F}_{M,\textsc{hom}}=N\int_{(\pi)}\dfrac{\D\vartheta}{\pi}\,\dfrac{1}{\sigma(\vartheta)^2}\dfrac{\partial\mu(\vartheta)}{\partial\rvec{a}}\dfrac{\partial\mu(\vartheta)}{\partial\rvec{a}}
\label{eq:Fhomm}
\end{equation}
after applying the central limit theorem, where $\mu(\vartheta)=\left<X^M_\vartheta\right>$, $\sigma(\vartheta)^2=\left<X^{2M}_\vartheta\right>-\left<X^M_\vartheta\right>^2$ and $\rvec{a}$ is the $(M+1)$-dimensional column of $M$th moment parameters. We primarily focus on $M=1$ and 2, for which the Fisher matrix in \eqref{eq:Fhomm} takes the forms
\begin{align}
\FHOMone&=N\int_{(\pi)}\dfrac{\D\vartheta}{\pi}\dfrac{\dyadic{m}_\vartheta}{\langle X^2_\vartheta\rangle-\langle X_\vartheta\rangle^2}\,,\nonumber\\
\FHOMtwo&=N\int_{(\pi)}\dfrac{\D\vartheta}{\pi}\dfrac{\dyadic{M}_\vartheta}{\langle X^4_\vartheta\rangle-\langle X^2_\vartheta\rangle^2}\,,
\label{eq:Fhomm12}
\end{align}
where $\rvec{m}_\vartheta=\rvec{u}_\vartheta\rvec{u}_\vartheta$, $\rvec{u}_\vartheta\,\widehat{=}\,\TP{(\cos\vartheta\,\,\,\sin\vartheta)}$ and
\begin{equation}
\dyadic{M}_\vartheta\,\,\widehat{=}\begin{pmatrix}
\left(\cos{\vartheta}\right)^2\\
\sqrt{2}\sin{\vartheta}\cos{\vartheta}\\
\left(\sin{\vartheta}\right)^2
\end{pmatrix}\begin{pmatrix}
\left(\cos{\vartheta}\right)^2 & \sqrt{2}\sin{\vartheta}\cos{\vartheta} & \left(\sin{\vartheta}\right)^2
\end{pmatrix}\,.
\end{equation}

The optimal estimators $\rest^{(\textsc{opt})}$ and $\Gest^{(\textsc{opt})}$ that go with the sCRBs can be constructed using the HOM data by adopting mathematical techniques developed in operator frame theory~\cite{Teo:2017aa}. The answers read
\begin{align}
\widehat{\rvec{r}}^{(\textsc{opt})}_\textsc{hom}&=\dyadic{W}_1^{-1}\sum^{n_\vartheta}_{k=1}\rvec{u}_k\dfrac{N_k\widehat{\left<X_k\right>}}{\widehat{\left<X^2_k\right>}-\widehat{\left<X_k\right>}^2}\,\nonumber\\
\dyadic{W}_1&=\sum^{n_\vartheta}_{k=1}\rvec{m}_{k}\dfrac{N_k}{\widehat{\left<X_k^2\right>}-\widehat{\left<X_k\right>}^2}\,,
\label{eq:opt1}
\end{align}
and
\begin{align}
\widehat{\dyadic{G}}^{(\textsc{opt})}_{2}&=\dyadic{W}_2^{-1}\sum^{n_\vartheta}_{k=1}\VEC{\dyadic{m}_k}\dfrac{N_k\widehat{\left<X_k^2\right>}}{\widehat{\left<X^4_k\right>}-\widehat{\left<X_k^2\right>}^2}\,,\nonumber\\
\dyadic{W}_2&=\sum^{n_\vartheta}_{k=1}\dyadic{M}_k\dfrac{N_k}{\widehat{\left<X_{k}^4\right>}-\widehat{\left<X_{k}^2\right>}^2}\,,
\label{eq:opt2}
\end{align}
where $N_k=\sum^{n_x}_{j=1}n_{jk}$ refers to the marginal sum of the binned data counts $n_{jk}$ for $n_\vartheta$ sampled phases $\{\vartheta_k\}$ and $n_x$ sampled voltage differences $\{x_j\}$ per phase, and finally the unbiased estimates of the operator moments are collectively defined by
\begin{equation}
\widehat{\left<X_k^M\right>}=\dfrac{1}{N_k}\sum^{n_x}_{j=1}n_{jk}x^M_{jk}\,.
\end{equation}
The vectorization $\VEC{\,\cdot\,}$ maps a $2\times2$ symmetric matrix to a $3\times1$ real column inasmuch as
\begin{equation}
\dyadic{Y}\,\widehat{=}\begin{pmatrix}
y_1 & y_2\\
y_2 & y_3
\end{pmatrix}\mapsto\VEC{\dyadic{Y}}\,\widehat{\equiv}\begin{pmatrix}
y_1\\
\sqrt{2}\,y_2\\
y_3
\end{pmatrix}\,.
\end{equation}

We remark that for all $M>1$, the integrals in \eqref{eq:Fhomm12} [or \eqref{eq:Fhomm} for that matter] have no known analytical form for arbitrary $\rho$. As shall be demonstrated, one can nonetheless obtain closed-form expressions for specific classes of quantum states.

\subsection{Heterodyning}
\label{subsec:het}

The simultaneous measurements of $X$ and $P$ [see Fig.~\ref{fig:Fig1}(b)] with HET realize the coherent-state measurement. This projects the state $\rho$ to coherent states \emph{via} a delocalized phase-space sampling of the Husimi function. It is easy to see that the additional vacuum noise introduced by the initial beam splitter physically infuses measurement uncertainty in the resulting $X-P$ measurement data of the Arthurs--Kelly type relation
\begin{equation}
\VAR{\textsc{q}}{x}\VAR{\textsc{q}}{p}=\left(\langle(\Delta X)^2\rangle+\frac{1}{2}\right)\left(\langle(\Delta P)^2\rangle+\frac{1}{2}\right)\geq1>\frac{1}{2}\,,
\end{equation}
which is saturated by coherent states $[\langle(\Delta X)^2\rangle=\langle(\Delta P)^2\rangle=1/2]$. The variance $\VAR{\textsc{q}}{y}=\overline{y^2}-\overline{y}^2$ is defined in terms of the Husimi-function average. In the phase-space language perspective, the action of the vacuum introduces an additive noise contribution to the covariance matrix $\dyadic{G}$,
\begin{equation}
\dyadic{G}_{\textsc{het}}=\dyadic{G}+\dfrac{\dyadic{1}}{2}\,.
\label{eq:Ghet_G_half}
\end{equation}
More generally, the switch from marginal sampling of the Wigner function to delocalized sampling of the Husimi function introduces higher-order noise to \emph{all} operator moment quantities (except for the first moments), a physical consequence of the Gauss--Weierstrass transform.

Since, any observable-parameter column, including a column of moments, can be expressed linearly in the Husimi function, we can define the unbiased estimators 
\begin{align}
\widehat{\rvec{r}}_\textsc{het}&\,\widehat{=}\,\dfrac{1}{N}\sum^N_{j=1}\begin{pmatrix}
x_j\\
p_j
\end{pmatrix}\,,\nonumber\\
\widehat{\dyadic{G}}_{2,\textsc{het}}&\,\widehat{=}\,\dfrac{1}{N}\sum^{N}_{j=1}\begin{pmatrix}
x_j^2 & x_jp_j\\
x_jp_j & p_j^2
\end{pmatrix}
\end{align}
that estimate the true quantities. Since both estimators are essentially sums of independent random variables, we may invoke the central limit theorem in the limit of large $N$ and conclude that the respective sCRBs are attained \emph{asymptotically} with these estimators for given HET data. In contrast with HOM, there exist formal analytical expressions for the HET sCRB for any $M$.

\section{First-moment tomography}
\label{sec:first-mom}

First moments are the only exception where analytical comparisons of HOM and HET are possible for any $\rho$. A direct evaluation of $\FHOMone$ and its inverse trace gives the concise expression
\begin{equation}
\sCRB_\textsc{1,hom}=\Tr{\dyadic{G}}+2\sqrt{\DET{\dyadic{G}}}\,.
\label{eq:CRB_HOM_first}
\end{equation}
On the other hand, the sCRB for HET is given by
\begin{align}
\sCRB_\textsc{1,het}&=\VAR{\textsc{q}}{x}+\VAR{\textsc{q}}{p}\nonumber\\
&=\Tr{\dyadic{G}_\textsc{het}}=\Tr{\dyadic{G}}+1\,.
\label{eq:CRB_HET_first}
\end{align}

It then follows immediately from the HRS uncertainty relation pointed out in Sec.~\ref{subsec:mom_power} that $\sCRB_{1,\textsc{het}}\leq\sCRB_{1,\textsc{hom}}$ for \emph{any} $\rho$. This implies that for \emph{all} quantum states, the tomographic power of HET is always greater than that of HOM in locating the average phase-space center of the quantum state. For minimum-uncertainty states, both schemes stand on equal footing ($\sCRB_{1,\textsc{hom}}=\sCRB_{1,\textsc{het}}$). This general result concludes the brief section.

\section{Second-moment tomography}
\label{sec:sec-mom}

For second-moment tomography, general analytical expressions for the sCRB are unavaliable. As such, the subsequent analysis on the performances of HOM and HET is carried out on individual classes of quantum states. For this purpose, we define $\gamma_2=\sCRB_{2,\textsc{het}}/\sCRB_{2,\textsc{hom}}$ to be the performance ratio that indicates the relative tomographic power between the two sampling schemes.

\subsection{Gaussian states}
\label{subsec:second_mom_gauss}

A detailed discussion on Gaussian states of nonzero $\rvec{r}$ is given in \cite{Teo:2017aa}. For the purpose of illustrating some principles, we restrict the present survey to Gaussian states of zero $\rvec{r}$. The corresponding covariance matrix
\begin{equation}
\dyadic{G}\,\,\widehat{=}\,\,\dyadic{O}\,\dfrac{\mu}{2}\begin{pmatrix}
\lambda & 0\\
0 & \frac{1}{\lambda}
\end{pmatrix}\TP{\dyadic{O}}\qquad\left(\dyadic{O}\TP{\dyadic{O}}=\dyadic{1}\right)
\end{equation}
for these centralized Gaussian states is effectively parametrized by the temperature $\mu\geq1$ that measures the thermality or size of the Gaussian uncertainty ellipse, and squeezing strength $\lambda\geq1$. For minimum-uncertainty states ($\mu=1$), we recover $\DET{\dyadic{G}}=1/4$. The complete expressions for the sCRBs are
\begin{align}
\sCRB_{2,\textsc{hom}}&=2\,\Tr{\dyadic{G}}\left(\Tr{\dyadic{G}}+3\sqrt{\DET{\dyadic{G}}}\right)\,,\nonumber\\
\sCRB_{2,\textsc{het}}&=2\left(\Tr{\dyadic{G}_\textsc{het}}^2-\DET{\dyadic{G}_\textsc{het}}\right)\,.
\end{align}

To simplify matters, we shall investigate two specialized forms of $\dyadic{G}=\dyadic{G}_2$. Let us first consider the case where $\lambda=1$, that is the class of thermal states. Since $\dyadic{G}$ is now a multiple of the identity, we get 
\begin{equation}
\gamma_2=\frac{3 (\mu +1)^2}{10 \mu ^2}
\end{equation}
for the second-moment performance ratio that is monotonically decreasing with $\mu$. For minimum-uncertainty states ($\mu=1$), $\gamma_2=6/5=1.2$, which is the maximum value. So HOM fares better than HET tomographically for Gaussian states near the vacuum state. As the Gaussian state becomes highly thermal ($\mu\rightarrow\infty$), we have $\gamma_2\rightarrow3/10$, which is the optimum value for \emph{all} centralized Gaussian states. The transition point occurs at $\mu=\frac{1}{7} \left(\sqrt{30}+3\right)\approx1.211$, which practically means that for almost all the thermal states, HET beats HOM, with higher significance for highly thermal states.

\begin{figure}[h!]
	\centering
	\includegraphics[width=0.7\columnwidth]{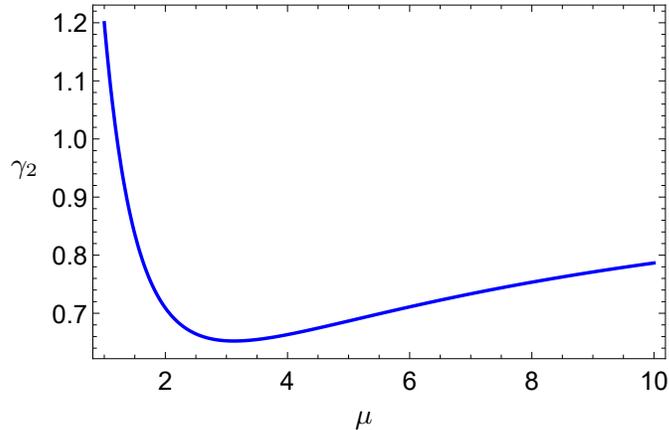}
	\caption{\label{fig:Fig2}A plot of $\gamma_2$ against $\mu=\lambda$ for centralized Gaussian states.}
\end{figure}

The other more interesting specialized case concerns the squeezed states ($\lambda>1$), where we shall just look at states with $\mu=\lambda$. These states approximately models strongly squeezed sources with excess noise associated to the anti-squeezed quadrature due to realistic experimental imperfections~\cite{Muller:2012ys}. In this case,
\begin{equation}
\gamma_2=\frac{\mu ^4+4 \mu ^2+7}{\left(\mu ^2+1\right) (\mu^2+3\mu+1)}\,.
\end{equation}
In the two extreme limits $\mu=1$ and $\mu\rightarrow\infty$, $\gamma_2$ takes the respective values $6/5$ and 1. This tells us that large squeezing ultimately reduces the tomographic benefits of HET over HOM, making the two techniques even for this specialized case. There exists an optimum $\gamma_2$ of $\approx0.652$ at $\mu\approx3.124$ (see Fig.~\ref{fig:Fig2}).

\subsection{Fock states}
\label{subsec:second_mom_fock}

\begin{figure}[h!]
	\centering
	\includegraphics[width=0.7\columnwidth]{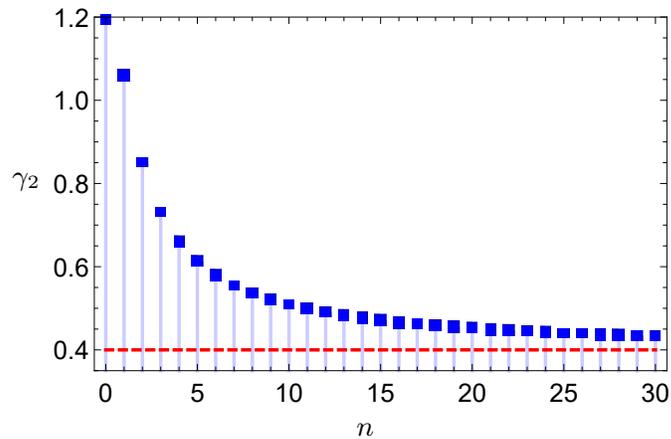}
	\caption{\label{fig:Fig3}A plot of $\gamma_2$ against $n$ for Fock states. The dashed red line marks the asymptotic value.}
\end{figure}

Fock states are important non-Gaussian quantum states that not only follow naturally from ideal conditions in photon-counting techniques, but are also crucial in the foundations of quantum mechanics. The covariance matrix for these states is always a multiple of the identity as they are spherically symmetric in phase space, so that we again arrive at the very simple formulas
\begin{align}
\sCRB_{2,\textsc{hom}}&=5\,(n^2+n+1)\,,\nonumber\\
\sCRB_{2,\textsc{het}}&=2\,(n+1)(n+3)\,,
\end{align}
and hence
\begin{equation}
\gamma_2=\dfrac{2\,(n+1)(n+3)}{5\,(n^2+n+1)}\,.
\end{equation}
For $n=0$, we evidently obtain the familiar answer $\gamma_2=6/5$ for the vacuum state, whereas for $n=1$, $\gamma_2=16/15$. In the limit of large $n$, $\gamma_2\rightarrow2/5$ (see Fig.~\ref{fig:Fig3}).

\subsection{Even and odd coherent states}
\label{subsec:second_mom_eo}

Another popular class of non-Gaussian states in continuous-variable quantum information theory with interesting phase-space quantum interference features are the even and odd coherent states defined by the ket $\ket{\pm;\alpha_0}=(\ket{\alpha_0}\pm\ket{-\alpha_0})\mathcal{N}_\pm$ with the normalization constant $\mathcal{N}_\pm=1/\sqrt{2\pm2\,\E{-2|\alpha_0|^2}}$. Without loss of generality, we may take $\alpha_0\geq0$~\cite{Teo:2017aa}.

For these slightly more sophisticated pure states, the matrix $\dyadic{F}_{2,\textsc{hom}}$ takes the form
\begin{align}
\dyadic{F}_{2,\textsc{hom}}&=\int_{(\pi)}\dfrac{\D\vartheta}{\pi}\,\dfrac{\dyadic{M}_\vartheta}{m_\pm+l\cos(2\vartheta)}\quad (l=2\alpha_0^2<m_\pm)\,,\nonumber\\
m_\pm&=\dfrac{1}{2}+2\alpha_0^2\left[\tanh\!\left(\alpha_0^2\right)\right]^{\pm1}\pm\dfrac{4\alpha_0^4}{\left(\E{\alpha_0^2}\pm\E{-\alpha_0^2}\right)^2}\,,
\end{align}
which leads to
\begin{equation}
\sCRB_{2,\textsc{hom}}=6m_\pm+4\sqrt{m_\pm^2-l^2}\,.
\label{eq:CRB_HOM_oe}
\end{equation}
The Husimi averaging of phase-space moments separately gives the sCRB
\begin{equation}
\sCRB_{2,\textsc{het}}=6+12\alpha_0^2\left[\tanh\!\left(\alpha_0^2\right)\right]^{\pm1}\pm\dfrac{8\alpha_0^4}{\left(\E{\alpha_0^2}\pm\E{-\alpha_0^2}\right)^2}\,.
\label{eq:CRB_HET_oe}
\end{equation}

\begin{figure}[h!]
	\centering
	\includegraphics[width=0.7\columnwidth]{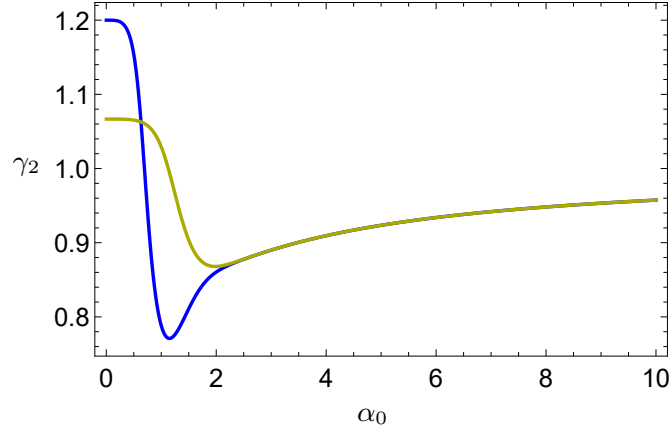}
	\caption{\label{fig:Fig4}A plot of $\gamma_2$ against $\alpha_0$ for both the even (blue) and odd (yellow) coherent states. The two curves have distinct values at $\alpha_0=0$.}
\end{figure}

Although the expression for $\gamma_2$ is now a nontrivial function of $\alpha_0$, the general behavior is similar to that of the Gaussian states (see Fig.~\ref{fig:Fig4}). Firstly, the respective limiting cases ($\alpha_0=0$) for the even and odd states coincide with the $n=0$ (6/5) and $n=1$ (16/15) Fock states, as they should be. For the even coherent states, the unit-$\gamma_2$ crossover occurs at $\alpha_0\approx0.693$, whereas for the odd coherent states, this happens at $\alpha_0\approx1.128$. Secondly, for each type of states, $\gamma_2$ possesses a stationary global minimum. For the even states, the minimum value of $\gamma_{2,\text{min}}=0.77096$ is attained at $\alpha_0=1.148\approx1$. For the odd states, this optimum value is $\gamma_{2,\text{min}}=0.86796$ and is achieved with $\alpha_0=1.980\approx2$.

\subsection{Displaced Fock and Photon-added coherent states}
\label{subsec:second_mom_dispfock_padd}

The displaced Fock states and the ``cousin'' states, namely the photon-added coherent states that differ only by a swap in the order of displacement and squeezing operations, represent two important classes that constitute the building blocks for understanding quasi-distributions~\cite{Wallentowitz:1996qo,Paris:1996do,Banaszek:1999qn} and quantum bosonic systems~\cite{Parigi:2004qc,Kim:2008bc,Zavatta:2009aa}.

The presence of two parameters for these states influences the complexity of the sCRB expressions, notably for the photon-added coherent states where no easy closed-form formulas are available. Detailed discussions for these two states have been done \cite{Teo:2017aa}. For this subsection, we shall consolidate the main physical results.

The two classes of states share common traits that are inherited from Fock states and Gaussian states. For $m>1$, $\gamma_2<1$ for all values of $\alpha_0$. Exceptions appear for $m=0,1$, where beyond certain threshold value of $\alpha_0$ that varies for different classes of states and $m$ value $\gamma_2$ transits from positive to negative values. For each $m$ value, there is a minimum stationary point for $\gamma_2$ and the functional dependence of the stationary point with $m$ can be systematically derived for the displaced Fock states, whereas for the photon-added coherent states, it is possible to perform curve fitting to deduce the asymptotic behavior for this stationary point.

\section{Two-mode states}
\label{sec:two-mode}

\subsection{General theory}

We begin the generalization of all previous discussions to multi-mode sources by first looking at two-mode states. Their first-moment column
\begin{equation}
\rvec{r}=\begin{pmatrix}
\rvec{r}_1\\
\rvec{r}_2
\end{pmatrix}\,,\quad\rvec{r}_l\,\widehat{=}\begin{pmatrix}
\langle X_l\rangle\\
\langle P_l\rangle
\end{pmatrix}\,,
\end{equation}
collects the two sets of single-mode first-moment expectation values, and their second-moment matrix,
\begin{align}
\dyadic{G}_2&=\begin{pmatrix}
\dyadic{A}_1 & \dyadic{A}_{12}\\
\TP{\dyadic{A}}_{12} & \dyadic{A}_2
\end{pmatrix}\geq-\dfrac{\I}{2}\begin{pmatrix}
\dyadic{\Omega} & \dyadic{0}\\
\dyadic{0} & \dyadic{\Omega}
\end{pmatrix}\,,\nonumber\\
\quad\dyadic{A}_l&\,\widehat{=}\begin{pmatrix}
\langle X^2_l\rangle & \!\!\!\!\!\!\frac{1}{2}\langle \{X_l,P_l\}\rangle\\
\frac{1}{2}\langle \{X_l,P_l\}\rangle & \langle P^2_l\rangle
\end{pmatrix}\,,\,\,\dyadic{A}_{12}\,\widehat{=}\begin{pmatrix}
\langle X_1X_2\rangle & \langle X_1P_2\rangle\\
\langle X_2P_1\rangle & \langle P_1P_2\rangle
\end{pmatrix}\,,
\end{align}
contains the positive single-mode terms $\dyadic{A}_l$s that separately obey the HRS inequality and the cross-mode term $\dyadic{A}_{12}$ that accounts for all two-mode correlations. We shall consider a natural situation where the five components $\rvec{r}_1$, $\rvec{r}_2$, $\dyadic{A}_1$, $\dyadic{A}_2$ and $\dyadic{A}_{12}$ are independently reconstructed with equal investments of the complete measurement data. Then, the sCRBs are just sums of those of the relevant independent components:
\begin{align}
\sCRB_{1}&=\sCRB^{(1)}_{1}+\sCRB^{(2)}_{1}\\
\sCRB_{2}&=\sCRB^{(1)}_{2}+\sCRB^{(2)}_{2}+\sCRB^{(12)}_{2}\,.
\end{align}

There are many ways to perform parameter reconstruction on two-mode sources. Here, we are interested in the straightforward extension from single- to two-mode measurement schemes through the tensor-product structure, where each optical mode is probed by the same scheme. After going through lengthy but straightforward statistical calculations, we would arrive at the expressions for the Fisher matrices with the product-HOM scheme. They are
\begin{align}
{\dyadic{F}_{m,\textsc{hom}}}^{(l)}=N\int_{(\pi)}\dfrac{\D\vartheta_l}{\pi}\dfrac{\dyadic{M}_{\vartheta_l}}{\left< X_{l,\vartheta_l}^{2m}\right>-\left< X_{l,\vartheta_l}^m\right>^2}
\end{align}
for the single-mode sectors with $l,m=1,2$, and
\begin{align}
{\FHOMtwo}^{(12)}=N\int_{(\pi)}\dfrac{\D\vartheta_1}{\pi}\int_{(\pi)}\dfrac{\D\vartheta_2}{\pi}\dfrac{\dyadic{m}_{\vartheta_1}\!\otimes\,\dyadic{m}_{\vartheta_2}}{\left< X_{1,\vartheta_1}^2X_{2,\vartheta_2}^2\right>-\left< X_{1,\vartheta_1}X_{2,\vartheta_2}\right>^2}
\end{align}
for the cross-mode sector. The formulas for the optimal estimators that reache the sCRBs defined by these Fisher matrix components can be derived accordingly.

The product-HET technique performs a delocalized sampling of the two-mode Husimi function. A simple adaptation of the arguments for the single-mode case in Sec.~\ref{subsec:het} allows us to conclude that the following estimators
\begin{equation}
\widehat{\rvec{r}}=\dfrac{1}{N}\sum^N_{j=1}\begin{pmatrix}
\widehat{\rvec{r}}_{1,j}\\
\widehat{\rvec{r}}_{2,j}
\end{pmatrix}\,,\quad\widehat{\rvec{r}}_{l,j}\,\widehat{=}\begin{pmatrix}
x_{l,j}\\
p_{l,j}
\end{pmatrix}\,,
\end{equation}
and
\begin{align}
\widehat{\dyadic{G}}_{2,\textsc{het}}&=\dfrac{1}{N}\sum^N_{j=1}\begin{pmatrix}
\widehat{\dyadic{A}}_{1,j} & \widehat{\dyadic{A}}_{12,j}\\
\TP{\widehat{\dyadic{A}}}_{12,j} & \widehat{\dyadic{A}}_{2,j}
\end{pmatrix}\,,\nonumber\\
\widehat{\dyadic{A}}_{l,j}&\,\widehat{=}\begin{pmatrix}
x^2_{l,j} & \!\!x_{l,j}p_{l,j}\\
x_{l,j}p_{l,j} & p^2_{l,j}
\end{pmatrix}\,,\,\,\widehat{\dyadic{A}}_{12,j}\,\widehat{=}\begin{pmatrix}
x_{1,j}x_{2,j} & x_{1,j}p_{2,j}\\
x_{2,j}p_{1,j} & p_{1,j}p_{2,j}
\end{pmatrix}\,,
\end{align}
are asymptotically optimal for $\rvec{r}$ and $\dyadic{G}_2$ in the limit of large $N$. With these estimators, the HET sCRB may be derived as
\begin{align}
\sCRB_{1,\textsc{het}}=&\,\VAR{\textsc{q}}{x_1}+\VAR{\textsc{q}}{p_1}+\VAR{\textsc{q}}{x_2}+\VAR{\textsc{q}}{p_2}\,,\nonumber\\
\sCRB_{2,\textsc{het}}=&\,\VAR{\textsc{q}}{x^2_1}+\VAR{\textsc{q}}{p^2_1}+2\VAR{\textsc{q}}{x_1p_1}\nonumber\\
&\,+\VAR{\textsc{q}}{x^2_2}+\VAR{\textsc{q}}{p^2_2}+2\VAR{\textsc{q}}{x_2p_2}\nonumber\\
&\,+2\VAR{\textsc{q}}{x_1x_2}+2\VAR{\textsc{q}}{x_1p_2}\nonumber\\
&\,+2\VAR{\textsc{q}}{x_2p_1}+2\VAR{\textsc{q}}{p_1p_2}\,,
\end{align}
where now $\textsc{q}$ refers to the two-mode Husimi function.

\subsection{First-moment tomography}

Just like the single-mode case, $\sCRB_{1,\textsc{hom}}$ takes the closed-form expression 
\begin{equation}
\mathcal{H}_{1,\textsc{hom}}=\Tr{\dyadic{G}^{(1)}+\dyadic{G}^{(2)}}+2\left(\sqrt{\DET{\dyadic{G}^{(1)}}}+\sqrt{\DET{\dyadic{G}^{(2)}}}\right)
\end{equation}
for any two-mode $\rho$, where
\begin{equation}
\dyadic{G}^{(l)}\,\widehat{=}\begin{pmatrix}
\langle\left(\Delta X_{l}\right)^2\rangle & \frac{1}{2}\left<\left\{\Delta X_{l},\Delta P_{l}\right\}\right>\\
\frac{1}{2}\left<\left\{\Delta X_{l},\Delta P_{l}\right\}\right> & \langle\left(\Delta P_{l}\right)^2\rangle
\end{pmatrix}\,.
\end{equation}

Since $\dyadic{G}^{(l)}$ also obeys the HRS uncertainty relation $\DET{\dyadic{G}^{(l)}}\geq1/4$, this again implies the universal inequality $\sCRB_{1,\textsc{hom}}\geq\sCRB_{1,\textsc{het}}$. Equality holds when both the marginalized $\dyadic{G}^{(1)}$ and $\dyadic{G}^{(2)}$ are respectively the covariance matrices of single-mode minimum-uncertainty states.

\subsection{Second-moment tomography---two-mode Fock states}

As the class of two-mode Fock states ($\ket{n_1}\ket{n_2}$) are product states, their two-mode expectation values are evidently products of single-mode expectation values. This allows us to easily obtain 
\begin{align}
\sCRB_{2,\textsc{hom}}&=5(n^2_1+n^2_2)+21(n_1+n_2)+18\,,\nonumber\\
\sCRB_{2,\textsc{het}}&=2(n^2_1+n^2_2)+16(n_1+n_2)+20\,.
\end{align}
It turns out that except for $\ket{0}\ket{0}$ ($n_1=n_2=0$) for which we have $\gamma_2=10/9>1$, product-HET still beats product-HOM for all other values of $n_1$ and $n_2$. The performance ratio $\gamma_2$ approaches the minimum value of 2/7 in the limit $n_1=n_2\rightarrow\infty$.

\subsection{Second-moment tomography---two-mode squeezed vacuum states}

The two-mode squeezed vacuum state of nonnegative squeezing parameter $\zeta\geq0$ is defined by
\begin{equation}
\ket{\textsc{sqv}}=\dfrac{1}{\cosh(\zeta)}\sum^\infty_{n=0}\ket{nn}\left[\tanh(\zeta)\right]^n\,.
\end{equation}
This two-mode entangled state is an important resource for many applications in continuous-variable quantum information theory~\cite{vLoock:1999aa,Ban:1999dc,Huang:2015aa,Ruo-Berchera:2015ci}.

We emphasize that although all first moments for this entangled state are zero, the joint first moments are not. More specifically, $\left<X_{1,\vartheta_1}X_{2,\vartheta_2}\right>=\sinh(\zeta)\,\cosh(\zeta)\,\cos(\vartheta_1+\vartheta_2)$ and $\overline{x_1x_2}=\overline{p_1p_2}=\sinh(\zeta)\cosh(\zeta)$. For the product-HOM scheme, the Fisher components can be calculated with the results
\begin{align}
\left<X^4_{l,\vartheta_l}\right>-\left<X^2_{l,\vartheta_l}\right>^2&=\dfrac{1}{2}\left[\cosh(2\zeta)\right]^2\,,\nonumber\\
\left<X^2_{1,\vartheta_1}X^2_{2,\vartheta_2}\right>-\left<X_{1,\vartheta_1}X_{2,\vartheta_2}\right>^2&=a\,\cos(2\vartheta_1+2\vartheta_2)+b\,,
\end{align}
where the coefficients $a=\dfrac{1}{2}\left[\sinh(2\zeta)\right]^2$ and $b=\dfrac{1}{4}\left[1+3\cosh(4\zeta)\right]$. For these states, the second-moment sCRBs still have closed-form expressions:
\begin{align}
\sCRB_{2,\textsc{hom}}&=\dfrac{11}{2}+\dfrac{13}{2}\cosh(4\zeta)+6\,\cosh(2\zeta)\sqrt{\cosh(4\zeta)}\nonumber\\
\sCRB_{2,\textsc{het}}&=4\left[\cosh(\zeta)\right]^2\left[2+3\cosh(2\zeta)\right]\,.
\end{align}
When $\zeta=0$, the sCRBs give the correct limiting values for the two-mode vacuum state. Lastly, we see that $\gamma_2$ monotonically decreases from $\gamma_2\Big|_{\zeta=0}=10/9$ all the way to the asymptotically optimum value $\gamma_{2,\text{min}}=6/(13+6\sqrt{2})\approx0.27926$. The transition point ($\gamma_2=1$) occurs at $\zeta\approx0.2063$.

\section{Conclusion}

When investigating the tomographic powers of different measurement schemes, care must be taken to avoid erroneous conclusions based on unfair or unsystematic comparisons. In this article, we analyzed the performances of balanced homodyne and heterodyne sampling schemes by considering a properly scaled and optimized accuracy measure that is well-known in statistics. We applied this study to various quantum states in moment tomography and showed that heterodyning can give significantly better reconstruction accuracies than balanced homodyning, which contradicts fabled tales of how the additional vacuum is overwhelmingly detrimental in parameter estimation problems.

\section*{Acknowledgments}

We acknowledge
financial support from the BK21 Plus Program (21A20131111123) funded by the Ministry of Education (MOE, Korea) and National Research Foundation of Korea (NRF), the NRF grant funded by the Korea government (MSIP) (Grant No. 2010-0018295), the Korea Institute of Science and Technology Institutional Program (Project No. 2E26680-16-P025), the European Research Council (Advanced Grant PACART), the Spanish MINECO (Grant FIS2015-67963-P), the Grant Agency of
the Czech Republic (Grant No. 15-03194S), and the IGA Project of the
Palack{\'y} University (Grant No. IGA PrF 2016-005).

\bibliographystyle{ws-ijqi}

\begin{thebibliography}{10}

\bibitem{Yuen:1983ba}
H.~P. Yuen and V.~W.~S. Chan, {\em Opt. Lett.} {\bf 8}  (1983) 177.

\bibitem{Abbas:1983ak}
G.~L. Abbas, V.~W.~S. Chan and T.~K. Yee, {\em Opt. Lett.} {\bf 8}  (1983) 419.

\bibitem{Schumaker:1984qm}
B.~L. Schumaker, {\em Opt. Lett.} {\bf 9}  (1984) 189.

\bibitem{Vogel:1989zr}
K.~Vogel and H.~Risken, {\em Phys. Rev. A} {\bf 40}  (1989) 2847.

\bibitem{Banaszek:1997ot}
K.~Banaszek and K.~W{\'o}dkiewicz, {\em Phys. Rev. A} {\bf 55}  (1997) 3117.

\bibitem{Arthurs:1965al}
E.~Arthurs and J.~L. Kelly, {\em Bell Syst. Tech. J.} {\bf 44}  (1965) 725.

\bibitem{Yuen:1982hh}
H.~P. Yuen, {\em Phys. Lett. A} {\bf 91}  (1982) 101.

\bibitem{Arthurs:1988aa}
E.~Arthurs and M.~S. Goodman, {\em Phys. Rev. Lett.} {\bf 60}  (1988) 2447.

\bibitem{Martens:1990al}
H.~Martens and W.~M. de~Muynck, {\em Found. Phys.} {\bf 20}  (1990) 357.

\bibitem{Martens:1991aa}
H.~Martens and W.~M. de~Muynck, {\em Phys. Lett. A} {\bf 157}  (1991) 441.

\bibitem{Raymer:1994aj}
M.~G. Raymer, {\em Am. J. Phys.} {\bf 62}  (1994) 986.

\bibitem{Trifonov:2001up}
A.~Trifonov, G.~Bj{\"o}rk and J.~S{\"o}derholm, {\em Phys. Rev. Lett.} {\bf 86}
   (2001) 4423.

\bibitem{Werner:2004as}
R.~F. Werner, {\em Quantum Info. Comput.} {\bf 4}  (2004) 546.

\bibitem{Rehacek:2015qp}
J.~{\v R}eh{\'a}{\v c}ek, Y.~S. Teo, Z.~Hradil and S.~Wallentowitz, {\em Sci.
  Rep.} {\bf 5}  (2015) p. 12289.

\bibitem{Muller:2016da}
C.~R. M\"{u}ller, C.~Peuntinger, T.~Dirmeier, I.~Khan, U.~Vogl, C.~Marquardt,
  G.~Leuchs, L.~L. S\'{a}nchez-Soto, Y.~S. Teo, Z.~Hradil and
  J.~\v{R}eh\'{a}\v{c}ek, {\em Phys. Rev. Lett.} {\bf 117}  (2016) p. 070801.

\bibitem{Croal:2016qs}
C.~Croal, C.~Peuntinger, B.~Heim, I.~Khan, C.~Marquardt, G.~Leuchs, P.~Wallden,
  E.~Andersson and N.~Korolkova, {\em Phys. Rev. Lett.} {\bf 117}  (2016) p.
  100503.

\bibitem{Teo:2017aa}
Y.~S. Teo, C.~R. M\"{u}ller, H.~Jeong, Z.~Hradil, J.~\v{R}eh\'{a}\v{c}ek and
  L.~L. S\'{a}nchez-Soto, {\em Phys. Rev. A} {\bf 95}  (2017) p. 042322.

\bibitem{Englert:QM2}
B.-G. Englert, {\em Lectures on {Q}uantum {M}echanics: Volume 2: Simple
  Systems} (World Scientific Publishing Co., 2006).

\bibitem{Angulo:1993aa}
J.~C. Angulo, {\em J. Phys. A: Math. Gen.} {\bf 26}  (1993) 6493.

\bibitem{Angulo:1994ps}
J.~C. Angulo, {\em Phys. Rev. A} {\bf 50}  (1994) 311.

\bibitem{Simon:9709030}
R.~Simon, M.~Selvadoray, Arvind and N.~Mukunda

\bibitem{Arvind:1998aa}
Arvind, N.~Mukunda and R.~Simon, {\em J. Phys. A: Math. Gen.} {\bf 31}  (1998)
  565.

\bibitem{Namiki:2012gs}
N.~R., {\em Phys. Rev. A} {\bf 85}  (2012) p. 062307.

\bibitem{Ivan:2012da}
J.~S. Ivan, N.~Mukunda and R.~Simon, {\em Quantum Inf. Process.} {\bf 11}
  (2012) 873.

\bibitem{Leverrier:2012qs}
A.~Leverrier, {\em Phys. Rev. A} {\bf 85}  (2012) p. 022339.

\bibitem{Thearle:2016aa}
O.~Thearle, S.~M. Assad and T.~Symul, {\em Phys. Rev. A} {\bf 93}  (2016) p.
  042343.

\bibitem{Muller:2012ys}
{\em New J. Phys.} {\bf 14}  (2012) p. 085002.

\bibitem{Wallentowitz:1996qo}
S.~Wallentowitz and W.~Vogel, {\em Phys. Rev. A} {\bf 53}  (1996) 4528.

\bibitem{Paris:1996do}
M.~G.~A. Paris, {\em Phys. Lett. A} {\bf 217}  (1996) 78.

\bibitem{Banaszek:1999qn}
K.~Banaszek and K.~W{\'o}dkiewicz, {\em Phys. Rev. Lett.} {\bf 82}  (1999)
  2009.

\bibitem{Parigi:2004qc}
V.~Parigi, A.~Zavatta, M.~S. Kim and M.~Bellini, {\em Science} {\bf 317}
  (2007) 1890.

\bibitem{Kim:2008bc}
M.~S. Kim, H.~Jeong, A.~Zavatta, V.~Parigi and M.~Bellini, {\em Phys. Rev.
  Lett.} {\bf 101}  (2008) p. 260401.

\bibitem{Zavatta:2009aa}
A.~Zavatta, V.~Parigi, M.~S. Kim, H.~Jeong and M.~Bellini, {\em Phys. Rev.
  Lett.} {\bf 103}  (2009) p. 140406.

\bibitem{vLoock:1999aa}
P.~van Loock and S.~L. Braunstein, {\em Phys. Rev. A} {\bf 61}  (1999) p.
  010302(R).

\bibitem{Ban:1999dc}
M.~Ban, {\em J. Opt. B: Quantum Semiclass. Opt.} {\bf 1}  (1999) L9.

\bibitem{Huang:2015aa}
K.~Huang, H.~Le~Jeannic, J.~Ruaudel, V.~Verma, M.~Shaw, F.~Marsili, S.~Nam,
  E.~Wu, H.~Zeng, Y.-C. Jeong, R.~Filip, O.~Morin and J.~Laurat, {\em Phys.
  Rev. Lett.} {\bf 115}  (2015) p. 023602.

\bibitem{Ruo-Berchera:2015ci}
I.~Ruo-Berchera, I.~P. Degiovanni, S.~Olivares, N.~Samantaray, P.~Traina and
  G.~M., {\em Phys. Rev. A} {\bf 92}  (2015) p. 053821.

\end{thebibliography}

\end{document}